\documentclass[aip,preprint,floatfix,amsfonts,amsmath,amssymb]{revtex4-1}

\usepackage{graphicx,dcolumn,bm}

\begin{document}

\title{Resonant Magnetization Switching Conditions of an Exchange-coupled Bilayer under Spin Wave Excitation}

\author{W. Zhou}
\email{syuu.z@imr.tohoku.ac.jp}
\affiliation{Institute for Materials Research, Tohoku University, Sendai 980-8577, Japan}

\author{T. Yamaji}
\affiliation{National Institute of Advanced Industrial Science and Technology, Tsukuba 305-8568, Japan}

\author{T. Seki}
\affiliation{Institute for Materials Research, Tohoku University, Sendai 980-8577, Japan}
\affiliation{JST PRESTO, Saitama 332-0012, Japan}
\affiliation{Center for Spintronics Research Network, Tohoku University, Sendai 980-8577, Japan}

\author{H. Imamura}
\affiliation{National Institute of Advanced Industrial Science and Technology, Tsukuba 305-8568, Japan}

\author{K. Takanashi}
\affiliation{Institute for Materials Research, Tohoku University, Sendai 980-8577, Japan}
\affiliation{Center for Spintronics Research Network, Tohoku University, Sendai 980-8577, Japan}

\date{\today}

\begin{abstract}
We systematically investigated the spin wave-assisted magnetization switching (SWAS) of a \textit{L}1$_0$-FePt / Ni$_{81}$Fe$_{19}$ (Permalloy; Py) exchange-coupled bilayer using a pulse-like rf field ($h_{\text{rf}}$) and mapped the switching events in the magnetic field ($H$) - $h_{\text{rf}}$ frequency ($f$) plane to reveal the switching conditions.
The switching occurred only in a limited region following the dispersion relationship of the perpendicular standing spin wave (PSSW) modes in Py.
The results indicate that SWAS is a resonant magnetization switching process, which is different from the conventional microwave assisted switching, and has the potential to be applied to selective switching for multilevel recording media.
\end{abstract}


\maketitle

\begin{figure}
\includegraphics{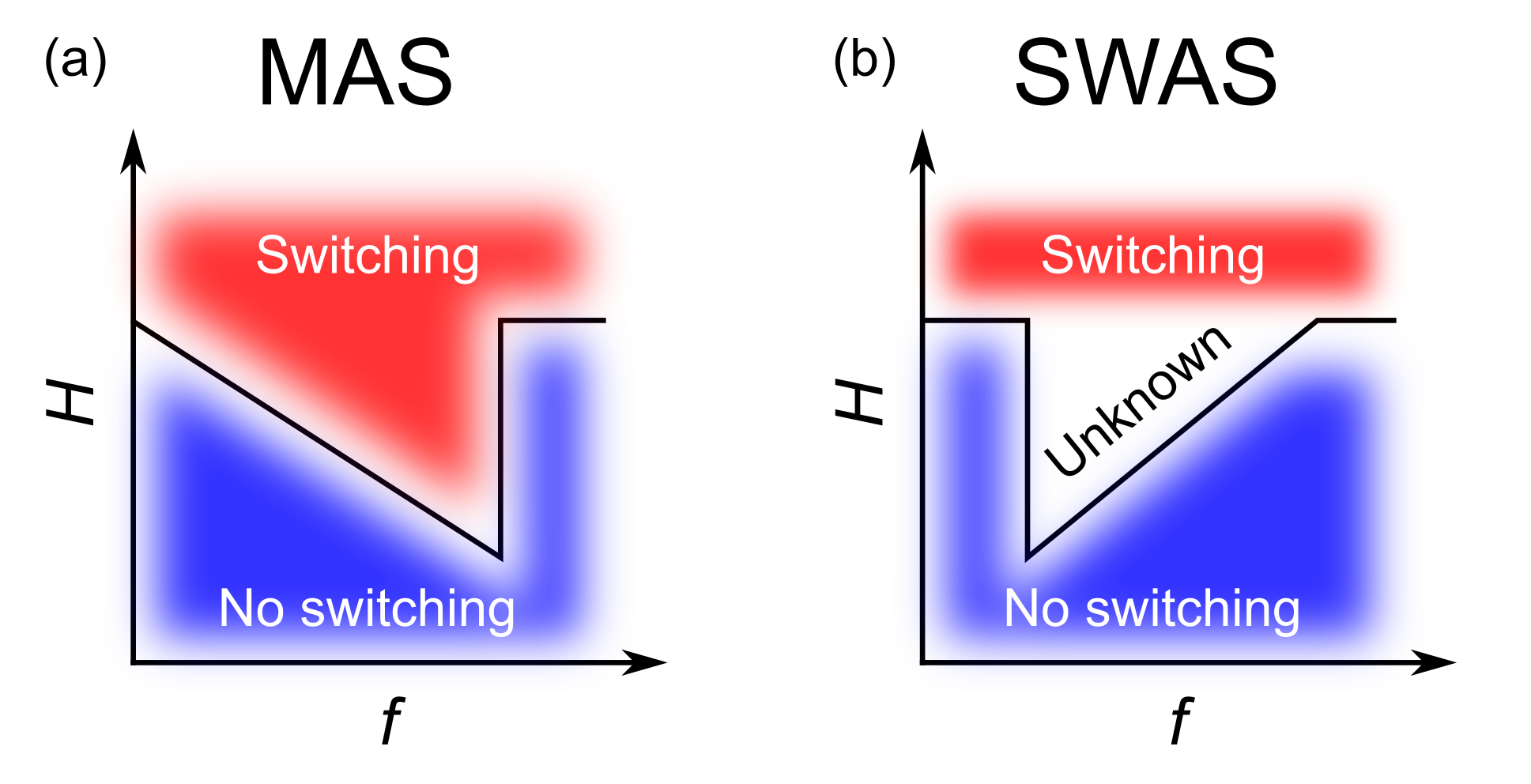}
\caption{\label{Fig1} Schematic illustrations of the switching conditions in the $H$ - $f$ plane for (a) microwave assisted switching (MAS), and (b) spin wave-assisted magnetization switching (SWAS). The areas of switching and no switching conditions are colored red and blue, respectively. The black lines indicates the variation of minimum $H_{\text{sw}}$ at different $f$.}
\end{figure}

In order to simultaneously fulfill the requirements of the thermal stability of magnetization and the writability of information for ultrahigh density magnetic storage devices, alternative recording technologies are desired to reduce the switching field ($H_{\text{sw}}$) of recording media materials with high magnetic anisotropy ($K_{\text{u}}$)\cite{HDD}.
Recently, a novel method called the spin wave-assisted magnetization switching (SWAS) has been proposed and demonstrated\cite{swas1,swas2,swas3,swas4,swas5}.
For this method, a static magnetic field ($H$) and a rf magnetic field ($h_{\text{rf}}$) are applied to the magnetic element, similar to the conventional microwave assisted switching (MAS)\cite{mas1,mas2,mas3,mas4,masSW,mas5,mas6,mas7,mas8}.
However, different from MAS, the element is an exchange-coupled bilayer consisting of a hard magnetic layer and a soft one instead of a single hard layer.
$h_{\text{rf}}$ excites the spin wave modes in the soft layer, which leads to the $H_{\text{sw}}$ reduction of the hard layer, whereas the behavior of MAS is closely related to the magnetization precession of the single hard layer.
As a result, the $h_{\text{rf}}$ frequency ($f$) dependence of $H_{\text{sw}}$ is completely different: for MAS, $H_{\text{sw}}$ decreases as $f$ increases till a critical $f$, where $H_{\text{sw}}$ reaches the minimum and then jumps back to the value without $h_{\text{rf}}$ (Fig.~\ref{Fig1}(a)); for SWAS, the minimum $H_{\text{sw}}$ is realized at a certain $f$, then $H_{\text{sw}}$ gradually increases as $f$ increases, and eventually return to the value without $h_{\text{rf}}$ application (Fig.~\ref{Fig1}(b)).
This grants SWAS the merit of large $H_{\text{sw}}$ reduction at relatively low $f$.
In addition, the three-dimensional magnetic recording using multilevel media has been intensively studied in recent years in order to further increase the recording density\cite{TD1,TD2,TD3,TD4,TD5}.
Realizing it requires a recording technique having the ability to selectively switch the magnetization of a desired layer without interfering the neighboring layers.
MAS has been exploited for this idea.
$H$ and $f$ are crucial parameters to decide the magnetization direction using MAS.
As shown in Fig.~\ref{Fig1}(a), the region above the $H_{\text{sw}}$ - $f$ curve corresponds to the switching conditions for MAS in the $H$ - $f$ plane\cite{mas3,mas6,mas7}.
As a result, in order to achieve selective switching in multilevel media using MAS, magnetic layers with different values of $K_{\text{u}}$ are used.
Each layer has a different $H_{\text{sw}}$ - $f$ curve, and the regions without overlapping with other curves give the selective switching conditions of each layer\cite{mas7,TD3,TD5}.
This indicates the importance of knowing the switching conditions in the $H$ - $f$ plane.
For SWAS, however, no systematic study has been conducted to provide such a map, and the switching condition in the $H$ - $f$ plane remains unclear as illustrated by the \textit{unknown} area in the middle of Fig.~\ref{Fig1}(b).
This information is also crucial when considering the application of SWAS for the three-dimensional magnetic recording.

In this letter, we report SWAS of a \textit{L}1$_0$-FePt / Ni$_{81}$Fe$_{19}$ (Permalloy; Py) exchange-coupled bilayer using a pulse-like $h_{\text{rf}}$.
The switching conditions in the $H$ - $f$ plane were revealed experimentally as well as through the numerical calculation.
The experimental and calculated results showed that the switching conditions of SWAS existed in a limited region in the  $H$ - $f$ plane, which followed the dispersion relationship of the perpendicular standing spin wave (PSSW) in Py.
The results clearly indicated a resonant magnetization switching process for SWAS.

\begin{figure}
\includegraphics{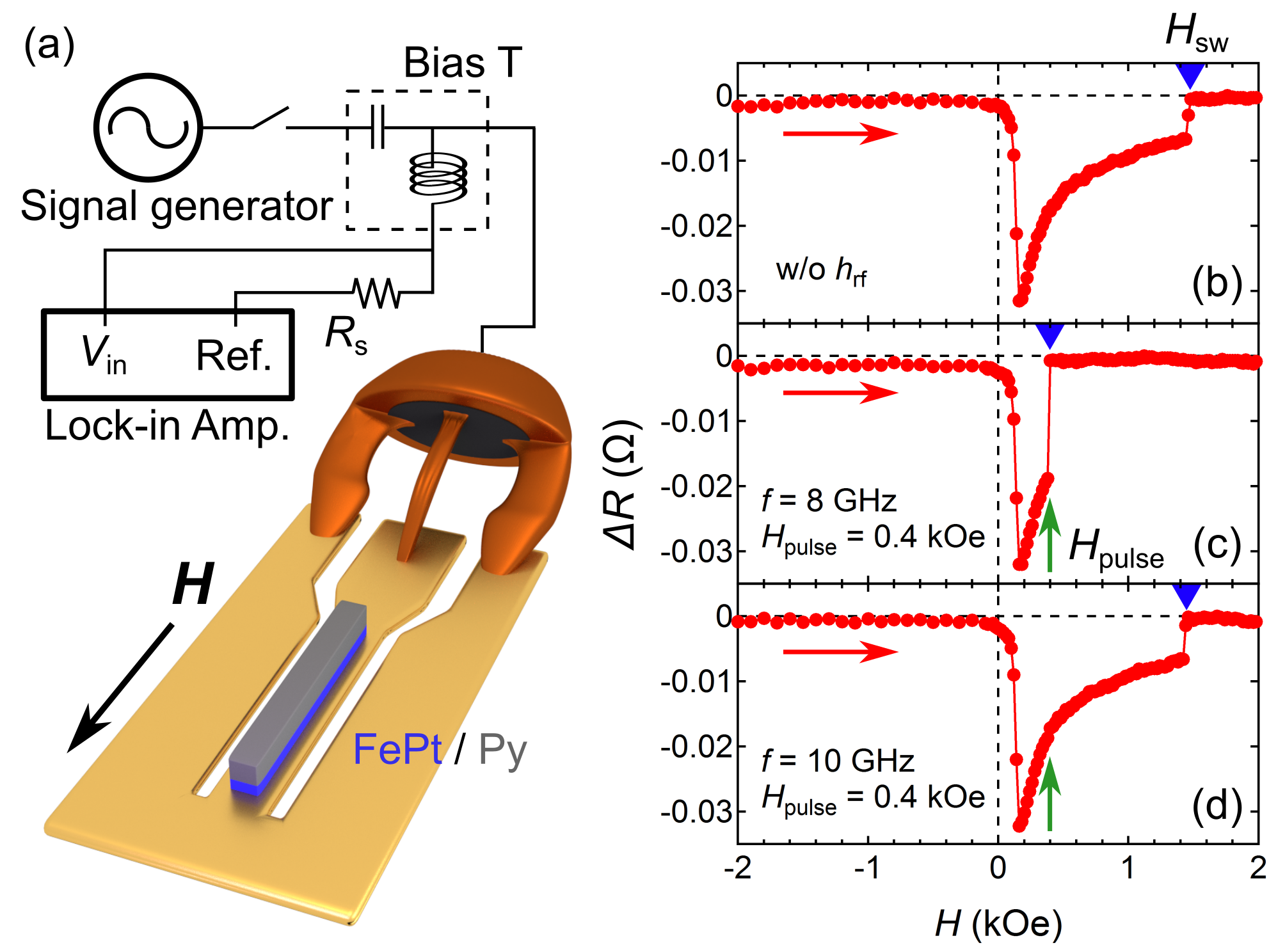}
\caption{\label{Fig2} (a) Schematic illustration of the device structure and the measurement setup. ${\Delta}R$ - $H$ curves measured (b) without $h_{\text{rf}}$ application, (c) with pulse-like $h_{\text{rf}}$ of 8 GHz applied at 0.4 kOe, (d) with pulse-like $h_{\text{rf}}$ of 10 GHz applied at 0.4 kOe. The red arrows indicate the sweeping directions of $H$, the blue triangles point out $H_{\text{sw}}$ of the FePt layer, and the green arrows mark the values of $H$ at which $h_{\text{rf}}$ was applied ($H_{\text{pulse}}$).}
\end{figure}

The exchange-coupled \textit{L}1$_{0}$-FePt / Py bilayer thin flim was prepared on an MgO (110) single crystal substrate.
The stacking structure of the thin film was MgO (110) subs. // Cr (10 nm) / Pt (10 nm) / Au (40 nm) / FePt (10 nm) / Py (100 nm) / Au (10 nm).
Layers from Cr to FePt were deposited using an ultrahigh vacuum magnetron sputtering system with the base pressure below $\sim 2\times10^{-7}$ Pa.
The Cr layer was deposited at room temperature (RT) then post-annealed at 600 $^{\circ}$C for 30 min.
The Pt layer was deposited at 300 $^{\circ}$C, followed by the deposition of the Au layer at RT.
Subsequently, the FePt layer was epitaxially grown on the Au layer at a substrate temperature of 350 $^{\circ}$C.
The substrate heating at 350 $^{\circ}$C promoted the formation of \textit{L}1$_{0}$-ordered structure.
The \textit{in-situ} reflection high energy electron diffraction observation was used to confirm the epitaxial growth of the FePt layer, which showed the epitaxial relation of MgO (110) $\parallel$ FePt (110), MgO [001] $\parallel$ FePt [001].
As a result, the FePt layer exhibited an in-plane uniaxial magnetic anisotropy along the MgO [001] direction\cite{FePt1,FePt2}.
Then, the sample was transferred to an ion beam sputtering system for the deposition of the Py layer and the Au capping layer at RT.
The compositions of FePt and Py were determined to be Fe$_{48}$Pt$_{52}$ and Ni$_{81}$Fe$_{19}$, respectively, by electron probe x-ray microanalysis.
Using the electron beam lithography and Ar ion milling, the FePt / Py bilayer was patterned into a rectangular element with the size of $2\times28$ $\mu$m$^2$, and the Cr/Pt/Au buffer layer was patterned into the shape of coplanar waveguide (CPW).
The schematic illustration of the device as well as the measurement setup is shown in Fig.~\ref{Fig2}(a).
In order to apply $h_{\text{rf}}$ to the device and to measure the device resistance ($R$) simultaneously, a lock-in amplifier and a standard resistance of 910 $\Omega$ were connected to the device through the DC port of a bias-Tee, while a signal generator (SG) was connected through the rf port.
The direction of $H$ was parallel to the in-plane MgO [001] direction, thus aligned with the in-plane easy axis of FePt.
The ferromagnetic resonance (FMR) spectra of the element were measured using a vector network analyzer (VNA)\cite{FMR}.
The numerical calculation was conducted by solving the Landau-Lifshitz-Gilbert (LLG) equation using the conventional fourth-order Runge-Kutta algorithm.
The model of an exchange-coupled FePt / Py bilayer with infinite size in the film plane was considered.
The thicknesses of FePt and Py were fixed at 10 nm and 100 nm, respectively.
The bilayer was discretized into 1-nm thick infinite plates with a uniform magnetization vector representing each plate.
The detail of the calculation can be found in Ref. \onlinecite{swas2}.
$K_{\text{u}}$ of FePt was set at $2.8\times10^{6}$ erg/cm$^{3}$, in order to reproduce the experimental value of $H_{\text{sw}}$ without $h_{\text{rf}}$ application.
The temperature was set at 300K throughout the calculation\cite{300K}.
For the switching events, the magnetization of the model was first aligned along the negative direction, then $H$ was applied.
After 10 ns to relax the system, $h_{\text{rf}}$ was applied for 10 ns then turned off, and another 10 ns was passed before the switching event of FePt was checked.

\begin{figure*}
\includegraphics{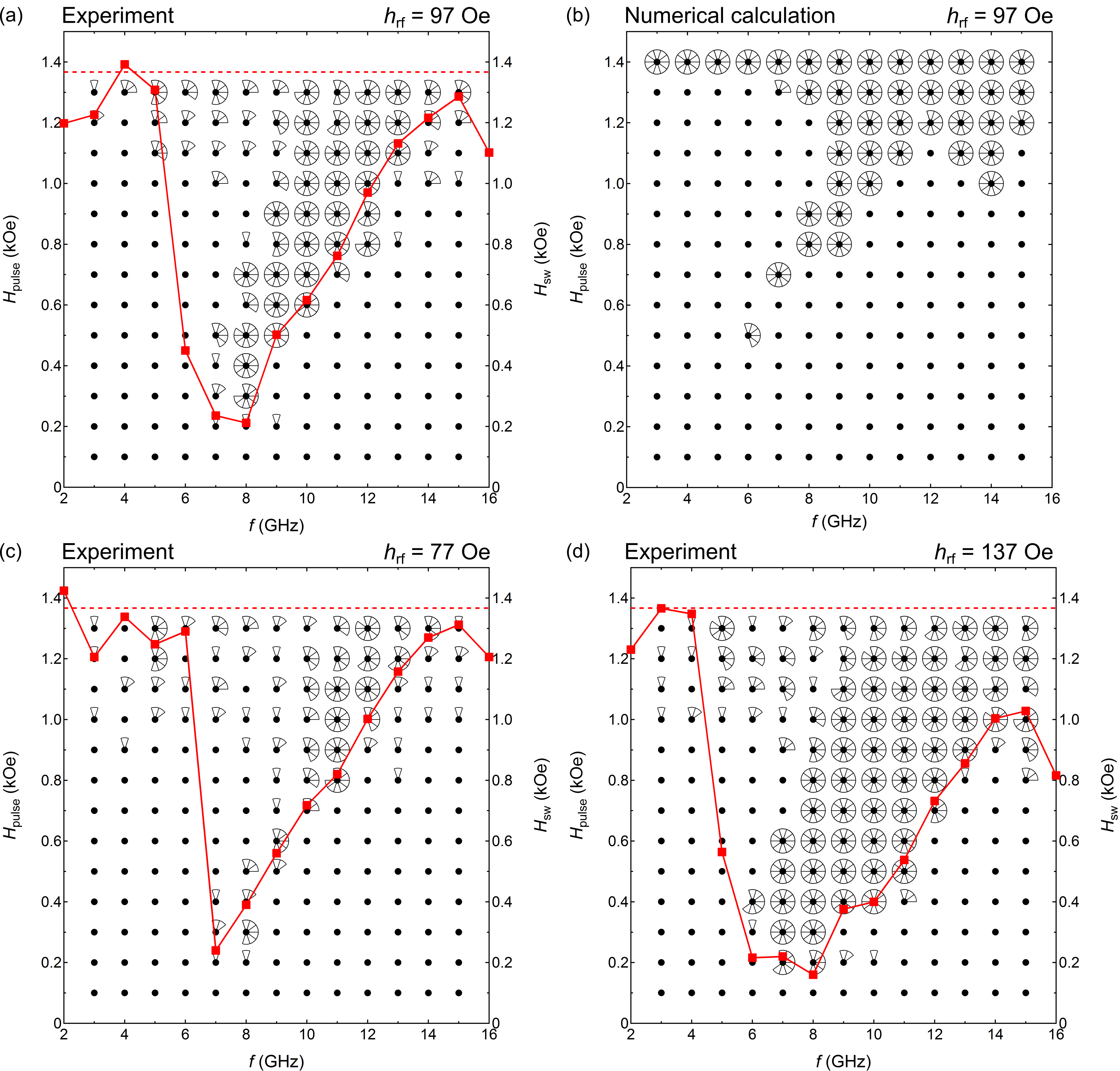}
\caption{\label{Fig3} (a) Experimentally mapped switching events of the FePt layer in the $H$ - $f$ plane. The rf power was set at 12 dBm, applying $h_{\text{rf}}$ of 97 Oe to the element. At each parameter in the plane, 10 measurements were carried out, and the number of switching events at  $H_{\text{pulse}}$ was represented by the number of sector. Red squares exhibit the averaged $H_{\text{sw}}$ measured under continuous $h_{\text{rf}}$ application with the same rf power input. The red dashed line indicates $H_{\text{sw}}$ without $h_{\text{rf}}$ application. (b) Switching conditions obtained from the numerical calculation. $h_{\text{rf}}$ was fixed at 97 Oe.(c) Experimentally mapped switching events of the FePt layer under $h_{\text{rf}}$ of 77 Oe (10 dBm) and (d) $h_{\text{rf}}$ of 137 Oe (15 dBm).}
\end{figure*}

$H_{\text{sw}}$ of the bilayer element was determined by measuring $R$.
${\Delta}R$ - $H$ curve without $h_{\text{rf}}$ application is shown in Fig.~\ref{Fig2}(b), where ${\Delta}R$ is the change of $R$ from the maximum value.
First, $H$ = $-$2 kOe was applied to the device to align all the magnetic moments of FePt and Py to the direction of $H$.
Then, $H$ was swept to 2 kOe while $R$ was monitored.
As $H$ became positive, the magnetic moments in Py rotated toward $H$ except the ones at the FePt / Py interface that were pinned by the adjacent FePt layer.
This led to the formation of the twisted state in Py.
${\Delta}R$ was mainly due to the anisotropic magnetoresistance (AMR) effect of Py.
In the twisted state, there were moments not parallel to the electrical current, resulted in a decrease of $R$\cite{AMR}.
As $H$ increased, the twisted spin structure was compressed towards the FePt / Py interface, and $R$ was gradually restored.
Finally, FePt was switched and $R$ jumped back to the maximum value.
Therefore, $H_{\text{sw}}$ of FePt was obtained from the ${\Delta}R$ - $H$ curve, as indicated by the blue triangle.
The ${\Delta}R$ - $H$ curve is consistent with previous reports\cite{swas2,swas3,swas4}.
In order to map the switching events of SWAS, the pulse-like $h_{\text{rf}}$ was applied to the device.
Unlike the method used previously, in which $h_{\text{rf}}$ was applied to the device continuously during the measurement of the ${\Delta}R$ - $H$ curves, SG fixed at an output of power and $f$ was turned on for 200 msec at a certain $H$ ($H_{\text{pulse}}$).
Figure~\ref{Fig2}(c) shows the ${\Delta}R$ - $H$ curve with $h_{\text{rf}}$ = 97 Oe (correspondent to the rf power of 12 dBm from SG) of 8 GHz applied at $H_{\text{pulse}}$ = 0.4 kOe.
${\Delta}R$ suddenly jumped to zero after the application of $h_{\text{rf}}$ indicated by the green arrow, thus $H_{\text{sw}}$ = $H_{\text{pulse}}$.
If $f$ was altered to 10 GHz (Fig.~\ref{Fig2}(d)), no clear change in ${\Delta}R$ was observed at $H_{\text{pulse}}$, and $H_{\text{sw}}$ similar to the case without $h_{\text{rf}}$ application was obtained.
Using this method, we are able to reveal the switching conditions in the $H$ - $f$ plane.

\begin{figure}
\includegraphics{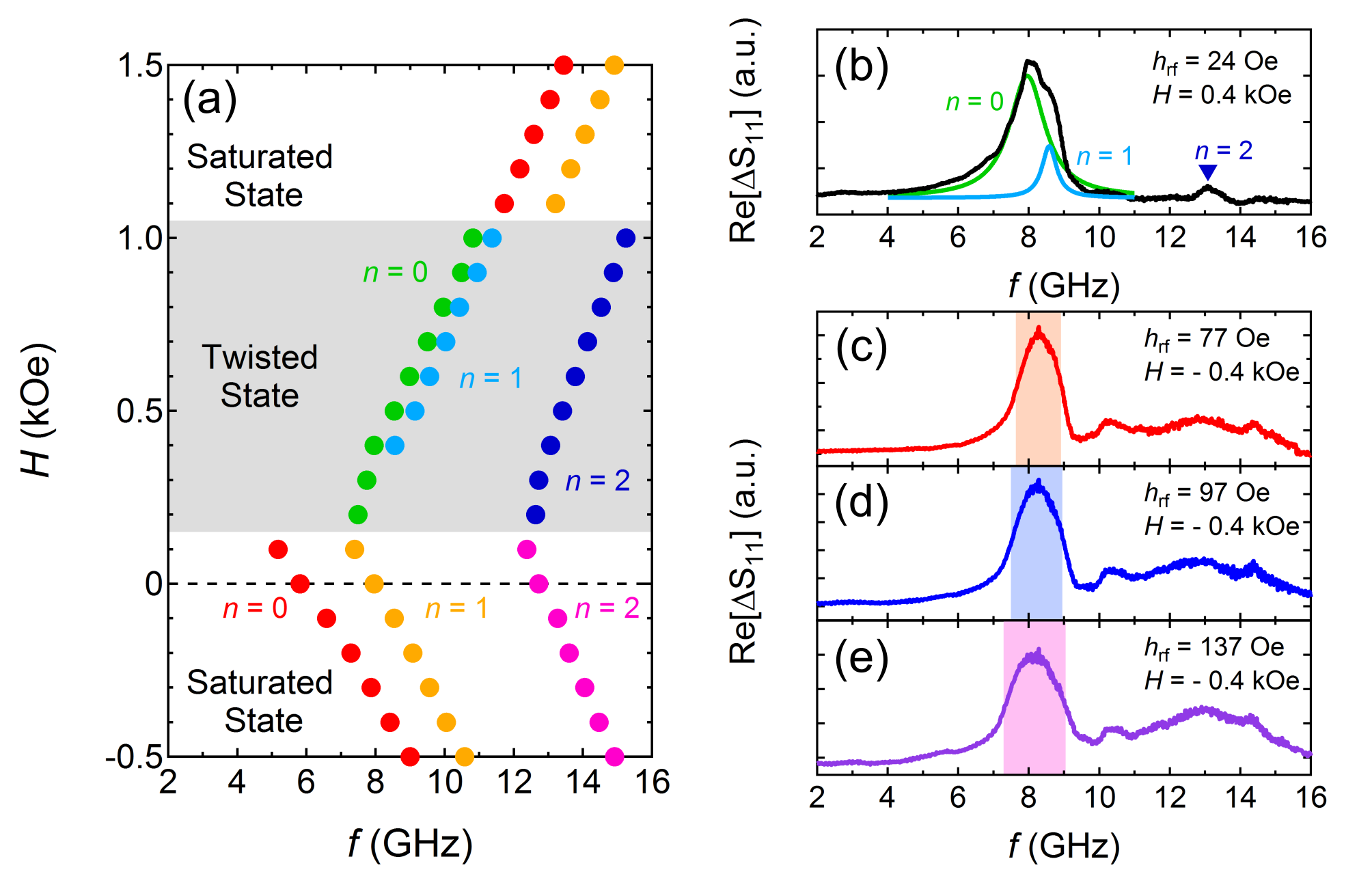}
\caption{\label{Fig4} (a) Dispersion relationship of the PSSW modes of the FePt / Py element in the $H$ - $f$ plane. The positions of the resonance peaks were obtained from the FMR spectra measured by VNA. (b) The FMR spectrum of the element at 0.4 kOe measured using $h_{\text{rf}}$ of 24 Oe, where the Py layer was in twisted state. The main peak was decomposed into two peaks by fitting to two Lorentz functions. The $n$ = 0 and 1 modes of PSSW were exhibited by the green and light blue curves, respectively. The $n$ = 2 mode was marked by the blue triangle. (c) The FMR spectra measured using $h_{\text{rf}}$ of 77 Oe, (d) 97 Oe and (e) 137 Oe at $-$0.4 kOe, where the Py layer was in saturated state. The colored areas indicates FWHM of the $n$ = 0 mode peaks of PSSW.}
\end{figure}

$f$ was varied from 3 GHz to 15 GHz with 1 GHz interval and $H_{\text{pulse}}$ was varied from 0.1 kOe to 1.3 kOe with 0.1 kOe interval, while the amplitude of $h_{\text{rf}}$ was fixed at 97 Oe.
At a certain $f$ and $H_{\text{pulse}}$, the ${\Delta}R$ - $H$ curves were measured ten times, and the results were summarized in Fig.~\ref{Fig3}(a).
If $f$ and $H_{\text{pulse}}$ satisfied the switching conditions, $H_{\text{pulse}}$ became $H_{\text{sw}}$, and one sector was added to represent it in the $H$ - $f$ plane.
As a result, a full circle (ten sectors) means ten switching events in ten measurements, and a black dot indicates no switching event occurred at $H_{\text{pulse}}$.
$H_{\text{sw}}$ was investigated also under the continuous application of $h_{\text{rf}}$, and the averaged $H_{\text{sw}}$ of ten measurements was plotted using red squares on the same figure to the y-axis on the right hand side.
(The y-axes on the left and right hand sides have different labels of $H_{\text{pulse}}$ and $H_{\text{sw}}$ but the same value of $H$.)
The average $H_{\text{sw}}$ without $h_{\text{rf}}$ is indicated by the red dashed line.
The switching events of SWAS existed mainly in a limited region above $H_{\text{sw}}$ under continuous $h_{\text{rf}}$.
Compared to the dispersion relationship of PSSW shown in Fig.~\ref{Fig4}(a), the limited region in the $H$ - $f$ plane agreed well with the $n$ = 0 and 1 modes of PSSW.
Because the $n$ = 0 and 1 modes were nearly overlapped (Fig.~\ref{Fig4}(b)), however, it was difficult to distinguish which mode mainly contributed to the switching in the $H$ - $f$ plane experimentally.
As $f$ increased, the switching events occurred at higher $H$ following the $H$ dependence of resonance frequency for PSSW modes.
At the region close to the $H_{\text{sw}}$ without $h_{\text{rf}}$ application, there were some switching events, which may be irrelevant to SWAS.
$H_{\text{sw}}$ without $h_{\text{rf}}$ application had a certain distribution, and the perturbation from $h_{\text{rf}}$ or the heat from rf power input might trigger the switching events without exciting the PSSW modes in Py.
One thing worth noticing here is that around $f\sim$ 8 GHz, there was an area at $H\sim$ 0.9 kOe showed no switching event.
The switching conditions obtained from numerical calculation is shown in Fig.~\ref{Fig3}(b), in which $h_{\text{rf}}$ was fixed at 97 Oe.
The calculated results qualitatively coincide with the experimental results, and the region of switching conditions at low $H$ due to SWAS was reproduced.
The reached minimum $H_{\text{sw}}$ was larger and the region was smaller than those of the experiment, which was due to the calculated model with infinite size in the film plane.
For the size of FePt used in the experiments, the switching events occurred through the process of reversed domain nucleation and propagation.
The edges of the element usually act as sites for reversed domain nucleation due to the effect, \textit{e.g.}, microfabrication damage\cite{FePt3}, which was not considered in the calculation.
The important fact obtained from calculation is that around $f\sim$ 8 GHz, there was also an area at $H\sim$ 1.1 kOe showing no switching event, similar to the experimental results.
This behavior was not observed in MAS\cite{mas3,mas6,mas7}.
The switching behavior of MAS has been analyzed using the Stoner-Wohlfarth (SW) model adopted in the rotating frame, and the SW astroid curve is modified into $H_{k}^{2/3}=\left(H+H_{\omega}\right)^{2/3}+h_{\text{rf}}^{2/3}$, where $H_{k}$ is the anisotropy field, $H_{\omega}=2{\pi}f/|{\gamma}|$ is the fictitious field in the rotating frame and $\gamma$ is the gyromagnetic ratio\cite{masSW}.
Based on this equation, the increase of $H$ will certainly satisfy the switching condition.
On the other hand, for SWAS the switching events do not occur when the spin wave is not excited, even under large $H$ in the opposite direction.
This serves as a strong evident for the resonant magnetization switching process of SWAS.
As a result, the switching conditions are confined in the area close to the dispersion relationship of spin wave in the $H$ - $f$ plane.

The switching conditions of the element were also mapped using different rf power, and the results with $h_{\text{rf}}$ of 77 Oe (10 dBm) and 137 Oe (15 dBm) are shown in  Figs.~\ref{Fig3}(c) and (d), respectively.
For the case of  $h_{\text{rf}}$ = 77 Oe, some switching events due to SWAS occurred at low $H$, which mostly followed the measured $H_{\text{sw}}$ curve under continuous $h_{\text{rf}}$ as well as the PSSW of the element (Fig.~\ref{Fig4}(a)).
However, few parameters showed a full circle (ten switching events in ten measurements) and the occurrences of SWAS at low $H$ seemed random.
This suggests that $h_{\text{rf}}$ = 77 Oe was near the threshold amplitude for SWAS.
In case of  $h_{\text{rf}}$ = 137 Oe,  a clear enlargement for the region of switching conditions was observed compared to the case of $h_{\text{rf}}$ = 97 Oe (Fig.~\ref{Fig3}(a)).
The region also followed the PSSW of the element.
The change of size of the region can be qualitatively explained using the FMR spectra of the element.
The FMR spectra of the element in saturated state ($H$ = $-$0.4 kOe) were measured using $h_{\text{rf}}$ of 77 Oe, 97 Oe and 137 Oe, and are shown in Figs.~\ref{Fig4}(c), (d) and (e), respectively.
The FMR spectra in the twisted state were not obtained under large $h_{\text{rf}}$ due to the instant occurrence of SWAS.
The colored belts in Figs.~\ref{Fig4}(c), (d) and (e) exhibited the full width at half maximum (FWHM) of the $n$ = 0 mode of PSSW.
As $h_{\text{rf}}$ increased, a obvious broadening of the resonance peak can be seen.
Since SWAS is caused by the excitation of the spin wave, the broadening of the resonance peak will lead to an enlargement of switching region in the $H$ - $f$ plane.
Meanwhile, the change of the amplitude of $h_{\text{rf}}$ did not cause a clear change to the minimum $H_{\text{sw}}$, which is quite different from the $h_{\text{rf}}$ amplitude dependence of MAS.
This behavior also indicates that SWAS is a resonant magnetization switching process.
It is worth pointing out that similar to the results in Fig.~\ref{Fig3}(a), the no switching area at higher $H$ above the switching region can also be observed under different amplitudes of $h_{\text{rf}}$ (Figs.~\ref{Fig3}(c) and (d)).

Based on the maps of the switching conditions, we consider SWAS to be capable of selective switching and suitable for three-dimensional magnetic recording.
The switching conditions of SWAS only existed in a limited region in the $H$ - $f$ plane following the dispersion relationship of spin wave, and the size of that region can be adjusted by the amplitude of $h_{\text{rf}}$.
Therefore, multiple FePt / Py bilayer units can be stacked in one element with their switching conditions densely packed but not overlapped by tuning the parameters of the bilayer, such as the thickness of the soft layer or $K_{\text{u}}$ of the hard layer or the strength of exchange-coupling.
The selective switching of one certain layer can be realized by application of $H$ and  $h_{\text{rf}}$ satisfying one's own switching conditions, similar to the idea exploited for MAS.
However, thanks to the limited region of switching conditions, the selective switching using SWAS may be achieved more easily with less variation of $H$ and $f$.
For instance, the aforementioned no switching area at the higher $H$ above the switching conditions in Figs.~\ref{Fig3} can be used as the switching conditions for another bilayer unit stacked in the same element, and the selective switching can be achieved by only changing $H$ without altering $f$.

In conclusion, we mapped the switching events of a \textit{L}1$_0$-FePt / Py exchange-coupled bilayer using a pulse-like $h_{\text{rf}}$ in both experiment and numerical calculation.
The results showed a limited region corresponds to the switching conditions of SWAS, which followed the dispersion relationship of the PSSW modes in Py.
As the amplitude of $h_{\text{rf}}$ increased, a broadening of the region was observed, which coincided with the broadening of the resonance peak of PSSW.
The results indicate that SWAS is a resonant magnetization switching process, and has potential for a selective switching technique in multilevel recording media.

\begin{acknowledgments}
This work was partially supported by Grant-in-Aid for Scientific Research B (16H04487). The device fabrication was partly performed at the Cooperative Research and Development Center for Advanced Materials, IMR, Tohoku University. 
\end{acknowledgments}

\bibliography{Text_ref}

\end{document}